\documentclass[runningheads]{llncs}
\usepackage[T1]{fontenc}
\usepackage{graphicx}
\usepackage{tabularx}
\usepackage{xcolor} 
\usepackage{booktabs}
\usepackage[normalem]{ulem} 
\usepackage{placeins}
\usepackage{amsmath}
\usepackage{amssymb}
\usepackage{multirow} 
\usepackage{amsmath}
\usepackage{orcidlink}
\begin{document}

\title{Aligning LLMs with Sparse Retrieval via Adaptive Sparse Supervision}

\author{
Tongna Chen\,\orcidlink{0009-0000-8585-669X}
\and
Minghan Li\,\orcidlink{0000-0002-1041-8887}
\thanks{Corresponding author: mhli@suda.edu.cn.}
\and
Guodong Zhou\,\orcidlink{0000-0002-7887-5099}
}

\authorrunning{T. Chen et al.}

\institute{
School of Computer Science and Technology,
Soochow University, Suzhou 215031, China\\
\email{tnchentnchen@stu.suda.edu.cn}\\
\email{mhli@suda.edu.cn}\\
\email{gdzhou@suda.edu.cn}
}

%
\maketitle              
\begin{abstract}
While large language models (LLMs) offer strong semantic representations, their application to first-stage retrieval incurs substantial latency and storage costs. Neural sparse retrievers such as SPLADE provide efficient retrieval but rely on relatively compact encoders with limited representational capacity. To bridge this gap, we propose an LLM-based sparse retrieval framework with input-adaptive sparse alignment. Specifically, we introduce a dynamic $\alpha$-truncation strategy that selects input-specific supervision supports based on the cumulative activation mass of SPLADE representations. This strategy guides the LLM to preserve salient lexical signals while suppressing less informative activations, without relying on a fixed support budget. Subsequent retrieval fine-tuning combines contrastive learning, knowledge distillation, and sparsity regularization to improve retrieval effectiveness. Experiments on MS MARCO and six selected BEIR datasets demonstrate improvements over SPLADE++ while maintaining competitive effectiveness against LLM-based sparse retrievers. Compared with Lion-sp, our method reduces index storage by 6.4× and end-to-end query latency by over 11×, achieving a favorable effectiveness–efficiency trade-off.

\end{abstract}
\section{Introduction}

Sparse retrieval remains central to information retrieval (IR), valued for its efficiency, interpretability, and compatibility with inverted indexing~\cite{zobel2006inverted,robertson2009probabilistic,bruch2024efficient}. 
Recent neural sparse retrievers such as SPLADE~\cite{formal2021splade,formal2022distillation}, uniCOIL~\cite{lin2021few}, and DeepImpact~\cite{mallia2021learning} extend this paradigm by learning high-dimensional vocabulary representations, achieving strong effectiveness while preserving the deployment advantages of classical sparse systems. 
However, these approaches typically rely on relatively small encoder architectures (e.g., BERT~\cite{devlin2019bert}), limiting their ability to fully exploit the semantic richness of large pre-trained language models (LLMs). In parallel, query expansion techniques based on pre-trained language models and LLMs have been extensively explored to alleviate lexical mismatch and enrich retrieval representations, as systematically reviewed by Li et al~\cite{10.1145/3816248}.

Large decoder-based LLMs offer substantially stronger semantic modeling capabilities, but integrating them into sparse retrieval remains challenging due to structural differences between continuous hidden representations and discrete lexical signals. 
Direct projection often leads to unstable sparsity and inefficient representations. 
This raises a key question: \emph{Can we retain LLM-level semantic expressiveness while preserving the efficiency and interpretability of sparse retrieval?}
\begin{figure}[t]
    \centering
    \includegraphics[width=\linewidth]{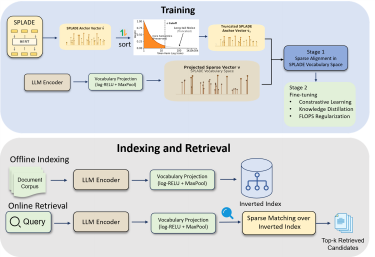}
    \caption{Overview of the proposed LLM-SPLADE framework. 
    Dense hidden states from the LLM are projected into the SPLADE vocabulary space, where input-adaptive $\alpha$-truncation preserves salient terms and suppresses long-tail noise. The resulting sparse representations are compatible with inverted indexing for efficient retrieval.}
    \label{fig:overview}
\end{figure}

Motivated by recent advances in neural sparse retrieval and the powerful semantic modeling capabilities of large language models, we investigate whether LLM-enhanced representations can be effectively integrated into sparse retrieval without sacrificing its inherent efficiency and interpretability. Drawing inspiration from the compatibility of sparse lexical systems with practical inverted indexing, we propose a lightweight alignment framework that bridges LLM semantic representations with high-dimensional sparse vocabulary spaces. Our key contributions are summarized as follows:

\begin{itemize}
    \item We introduce an input-adaptive $\alpha$-truncation mechanism that dynamically controls sparsity based on semantic distribution, which mitigates structural mismatch, suppresses long-tail noise, and preserves dominant semantic signals without task-specific fine-tuning.

    \item Extensive experiments demonstrate that our method improves in-domain retrieval performance over strong BERT-based sparse baselines and maintains competitive out-of-domain effectiveness under stricter sparsity constraints.

    \item Compared with prior LLM-based sparse retrievers, our approach achieves an average sparsity of $L_{0}=182$, reducing index storage by $6.4\times$ and improving query-time latency by over $11\times$, showcasing a favorable effectiveness--efficiency trade-off compatible with practical inverted-index deployment.
\end{itemize}

\section{Related Work}

\textbf{Neural Sparse Retrieval.}
Learned sparse retrievers such as DeepImpact~\cite{mallia2021learning}, uniCOIL~\cite{lin2021few}, and SPLADE~\cite{formal2021splade,formal2022distillation} improve classical lexical matching by learning term importance within the vocabulary space. 
These models retain compatibility with inverted indices while achieving substantial gains over traditional methods like BM25.

\textbf{Dense Retrieval.}
Dense retrieval methods leverage pre-trained language models to encode queries and documents into continuous embeddings~\cite{karpukhin2020dense,izacard2021unsupervised,hofstatter2021efficiently}. Beyond representation learning, self-supervised domain adaptation through pseudo-relevance labeling has also been explored to improve dense and conversational dense retrieval across domains ~\cite{li-gaussier-2024-domain}.
While effective at modeling semantic similarity, they require approximate nearest neighbor search for scalability~\cite{johnson2019billion}, and sacrifice the transparency and efficiency of lexical inverted indexing.

\textbf{LLM-based Retrieval.}
Recent work explores scaling decoder-only LLMs for sparse retrieval, demonstrating strong effectiveness and zero-shot generalization~\cite{zeng2025scaling,doshi2024mistral}. 
Another line of work studies how to align neural representations with interpretable lexical spaces. For example, MILCO~\cite{nguyen2026iclr-milco} introduces a multilingual connector that maps queries and documents into a shared English lexical space for multilingual and cross-lingual learned sparse retrieval. While these methods highlight the importance of lexical-space alignment, existing LLM-based sparse retrievers often produce highly dense activations, resulting in substantial storage and latency overhead. Another line of research explores LLM-based query expansion to enhance lexical matching and retrieval effectiveness. Li et al ~\cite{10.1145/3816248} provide a comprehensive survey of query expansion techniques in the PLM and LLM era. Recent studies further investigate prompt-based expansion and rank fusion ~\cite{LI2026104160}, as well as automatic exemplar construction and multi-LLM expansion refinement ~\cite{li2026automaticindomainexemplarconstruction}. Unlike these generation-oriented approaches, our method directly learns sparse vocabulary representations through LLM-to-SPLADE alignment, without explicitly generating expansion text. Our work focuses on aligning LLM representations with the SPLADE vocabulary space while explicitly controlling sparsity through input-adaptive $\alpha$-truncation.

\section{Method}

Our goal is to align dense representations produced by a pre-trained language model with SPLADE’s high-dimensional sparse vocabulary space, enabling efficient sparse retrieval while leveraging the semantic richness of LLM representations.
To bridge the structural gap between dense hidden states and sparse term-based representations, inspired by the Scaling-Retriever paradigm~\cite{zeng2025scaling}, we propose a two-stage framework:
(1) project encoder outputs into the vocabulary space with sparse-compatible activations, and
(2) enforce adaptive, input-aware sparsity via a dynamic $\alpha$-truncation mechanism.
Following common practices in large-scale retriever models, we adopt a bidirectional attention configuration~\cite{zeng2025scaling} to better capture contextual dependencies in the encoder representations.

\subsection{High-Dimensional Projection}

Given an input sequence, a pre-trained encoder produces hidden states $\mathbf{h} \in \mathbb{R}^{L \times d}$. We project these states into the vocabulary dimension $V$ and obtain the final sparse representation $\vec{v}$ via max-pooling~\cite{formal2022distillation}:

\begin{equation}
\vec{v} = \log \left( 1 + \text{ReLU} \left( \text{MaxPool} \left( \mathbf{h}\mathbf{W}_p + \mathbf{b}_p \right) \right) \right) \in \mathbb{R}^V,
\end{equation}

where $\mathbf{W}_p \in \mathbb{R}^{d \times V}$ and $\mathbf{b}_p \in \mathbb{R}^{V}$ are learnable parameters, and the $\log(1+\text{ReLU}(\cdot))$ transformation ensures non-negativity and sparsity.
\subsection{Adaptive $\alpha$-Truncation for Sparse Alignment}

Let $\hat{\mathbf{v}} \in \mathbb{R}^{V}$ denote a reference sparse vector from SPLADE, treated as a structured anchor space rather than a teacher. 
We define a dynamic truncation mask that adapts to the input’s activation distribution.
Let $\pi$ be the permutation of indices that sorts $\hat{\mathbf{v}}$ in descending order,
and select the smallest set of top-ranked indices whose cumulative mass exceeds a fraction $\alpha$ of the total:
\begin{equation}
K = \min \left\{ k \; \middle| \;
\sum_{j=1}^{k} \hat{v}_{\pi_j} \ge \alpha \sum_{i=1}^{V} \hat{v}_i
\right\}, \quad
\mathcal{T}_{\alpha} = \{ \pi_1, \ldots, \pi_K \}.
\end{equation}
This yields a content-aware, input-specific sparsity pattern that adapts to the distribution of dominant terms.
Conceptually, this formulation is related to mass-based sparse vector selection, such as the $\alpha$-mass subvector used in efficient retrieval over learned sparse representations~\cite{bruch2024efficient}. 
However, our goal is different: instead of organizing an index for approximate retrieval, we use the cumulative activation mass of the SPLADE anchor to construct an input-specific supervision support for aligning LLM representations with the sparse vocabulary space.

The model is trained to align with this structured support, while suppressing non-critical activations:
\begin{equation}
\mathcal{L}_{\text{align}} =
\frac{1}{|\mathcal{T}_\alpha|}
\sum_{i \in \mathcal{T}_\alpha} (v_i - \hat{v}_i)^2
+
\lambda \sum_{i \notin \mathcal{T}_\alpha} v_i^2.
\end{equation}
This objective encourages accurate approximation of the key active terms while maintaining sparsity. In practice, we optimize this objective via LoRA-based fine-tuning~\cite{hu2022lora}, which efficiently adapts the pre-trained encoder to the proposed alignment objective while keeping the majority of model parameters fixed.

\subsection{Contrastive Fine-Tuning with Knowledge Distillation and Sparsity Regularization}

After stage-one sparse alignment, we further fine-tune the model with LoRA-based lightweight adaptation. The stage-two objective combines contrastive learning, KL-based knowledge distillation, and FLOPS-style~\cite{formal2021splade} sparsity regularization to improve retrieval effectiveness while maintaining sparse and index-compatible representations. In our experiments, the teacher relevance scores used for KL-based distillation are obtained from MiniLM-cross-encoder-msmarco.

Given a query representation $\mathbf{v}_q$, a positive document representation $\mathbf{v}_d^+$, and a set of negative document representations $\mathcal{N}_q$, we first use a standard contrastive objective:
\begin{equation}
\mathcal{L}_{\mathrm{CL}}
=
-\log
\frac{
\exp\left(s(\mathbf{v}_q,\mathbf{v}_d^+)\right)
}{
\exp\left(s(\mathbf{v}_q,\mathbf{v}_d^+)\right)
+
\sum_{\mathbf{v}_d^- \in \mathcal{N}_q}
\exp\left(s(\mathbf{v}_q,\mathbf{v}_d^-)\right)
}.
\end{equation}
where $s(\cdot,\cdot)$ denotes the relevance score.

For knowledge distillation, we align the student ranking distribution with the teacher ranking distribution over the same candidate set. For each query $q$, let $\mathcal{D}_q = \{d^+\} \cup \{d_i^-\}_{i=1}^{N}$ denote the candidate documents. The student and teacher distributions are defined as:
\begin{equation}
P_{\mathrm{S}}(d \mid q)
=
\frac{
\exp\left(s_{\mathrm{S}}(q,d)/\tau\right)
}{
\sum_{d' \in \mathcal{D}_q}
\exp\left(s_{\mathrm{S}}(q,d')/\tau\right)
},
\end{equation}
\begin{equation}
P_{\mathrm{T}}(d \mid q)
=
\frac{
\exp\left(s_{\mathrm{T}}(q,d)/\tau\right)
}{
\sum_{d' \in \mathcal{D}_q}
\exp\left(s_{\mathrm{T}}(q,d')/\tau\right)
},
\end{equation}
where $s_{\mathrm{S}}(\cdot,\cdot)$ and $s_{\mathrm{T}}(\cdot,\cdot)$ denote the relevance scores produced by the student and teacher models, respectively, and $\tau$ is the temperature parameter.

The KL-based distillation loss is defined as:
\begin{equation}
\mathcal{L}_{\mathrm{KD}}
=
\frac{1}{|\mathcal{B}|}
\sum_{q \in \mathcal{B}}
\sum_{d \in \mathcal{D}_q}
P_{\mathrm{T}}(d \mid q)
\log
\frac{
P_{\mathrm{T}}(d \mid q)
}{
P_{\mathrm{S}}(d \mid q)
}.
\end{equation}

To preserve sparsity during fine-tuning, we adopt FLOPS-style regularization for both query and document representations. The regularization term is defined as:
\begin{equation}
\mathcal{L}_{\mathrm{FLOPS}}^{x}
=
\frac{1}{V}
\sum_{i=1}^{V}
\left(
\frac{1}{B_x}
\sum_{j=1}^{B_x}
v_{ij}^{x}
\right)^2,
\quad x \in \{q,d\},
\end{equation}
where $V$ is the vocabulary size, $B_x$ is the batch size, and $v_{ij}^{x}$ denotes the activation of vocabulary term $i$ in the $j$-th query or document representation.

The final stage-two objective is:
\begin{equation}
\mathcal{L}_{\mathrm{stage2}}
=
\frac{1}{2}
\left(
\mathcal{L}_{\mathrm{CL}}
+
\mathcal{L}_{\mathrm{KD}}
\right)
+
\lambda_q \mathcal{L}_{\mathrm{FLOPS}}^{q}
+
\lambda_d \mathcal{L}_{\mathrm{FLOPS}}^{d},
\end{equation}
where $\lambda_q = 0.05$ and $\lambda_d = 0.04$ control the sparsity of query and document representations, respectively.

This objective integrates contrastive learning and knowledge distillation with separate sparsity constraints on query and document representations, encouraging efficient inverted-index retrieval while maintaining retrieval effectiveness.

\section{Experiments}

\noindent \textbf{Hardware Setup.} 
Experiments were run on a workstation with Dual Intel Xeon Gold 6248R CPUs and a single NVIDIA A100 GPU (40GB). Query encoding (LLM inference) used the GPU, while sparse retrieval was performed on CPU with Pyserini and AVX-512.

\noindent \textbf{Training Details.} 
Stage~2 training uses a learning rate of $1\mathrm{e}{-4}$, with task weights $[1.0, 0.05, 0.04]$ corresponding to ranking, query regularization, and document regularization objectives. 
We use 16 negative samples per query and LoRA with rank $r=16$. 
Training is performed for 1 epoch with a batch size of 28 and gradient accumulation over 4 steps, totaling approximately 4700 optimization steps per epoch.

\subsection{Analysis of Truncation Strategies}
We adopt LLaMA-3.2-1B~\cite{dubey2024llama} with two-stage training: Stage~1 performs sparse alignment pre-training on in-domain data, and retrieval ablations are run on a 222\,MB MS\,MARCO subset, monitoring query (Top-10) and document (Top-20) gradient statistics. To assess input-adaptive support selection, we compare our dynamic $\alpha$-truncation with three fixed Top-$K$ baselines ($\alpha$-min, $\alpha$-mean, $\alpha$-max), whose $K$ values are derived from the minimum, mean, and maximum observed support sizes under the cumulative-mass criterion across training samples.

\textit{Top-Ratio} measures the proportion of gradient magnitude concentrated on the top-$K$ SPLADE-ranked terms, while \textit{Gradient Entropy} reflects the dispersion of gradients across the vocabulary. For SPLADE++, these statistics are computed from ranked sparse output weights as a proxy for gradient-based analysis.

\noindent \textbf{Mechanism Analysis.}
Table~\ref{tab:retrieval} reports the retrieval performance and gradient statistics of different truncation strategies, with $q_\alpha=0.8$ and $d_\alpha=0.8$ for dynamic $\alpha$-truncation.
$\alpha$-min uses the smallest observed support size, resulting in highly concentrated gradients (Entropy $1.70$, Top-Ratio $0.99$).
$\alpha$-max uses the largest support size, retaining more long-tail activations (Entropy $3.20$, Top-Ratio $0.49$).
$\alpha$-mean adopts a fixed average support budget but lacks input adaptivity (Entropy $2.95$, Top-Ratio $0.61$).
In contrast, dynamic $\alpha$-truncation selects an input-specific support size, achieving better retrieval performance than the fixed-budget alternatives while maintaining balanced gradient statistics (Entropy $2.92$, Top-Ratio $0.67$).

\begin{table}[t]
\centering
\small
\setlength{\tabcolsep}{4pt}
\caption{Retrieval performance (MRR@10) and gradient statistics. Top-Ratio and Grad Entropy are computed for query (Q) and document (D).}
\label{tab:retrieval}
\begin{tabular}{l c cc cc}
\toprule

\multirow{2}{*}{\textbf{Method}} & \multirow{2}{*}{\textbf{MRR@10}} & \multicolumn{2}{c}{\textbf{Top-Ratio}} & \multicolumn{2}{c}{\textbf{Grad Entropy}} \\
\cmidrule(lr){3-4} \cmidrule(lr){5-6}
 & & Q & D & Q & D \\
\midrule
SPLADE++ (Baseline) & 0.352 & 0.68 & 0.73 & 3.12 & 3.00 \\
\midrule
\multicolumn{6}{l}{\textit{LLaMA-1B w/ different strategies}} \\
~~+ $\alpha$-min & 0.069 & 0.99 & 0.89 & 1.70 & 2.93 \\
~~+ $\alpha$-mean & 0.303 & 0.61 & 0.44 & 2.95 & 3.89 \\
~~+ $\alpha$-trunc (Ours) & \textbf{0.322} & 0.67 & 0.47 & 2.92 & 3.91 \\
~~+ $\alpha$-max & 0.286 & 0.49 & 0.37 & 3.20 & 4.14 \\
\bottomrule
\end{tabular}
\end{table}

\begin{figure}[t] 
    \centering 
    \includegraphics[width=0.9\linewidth]{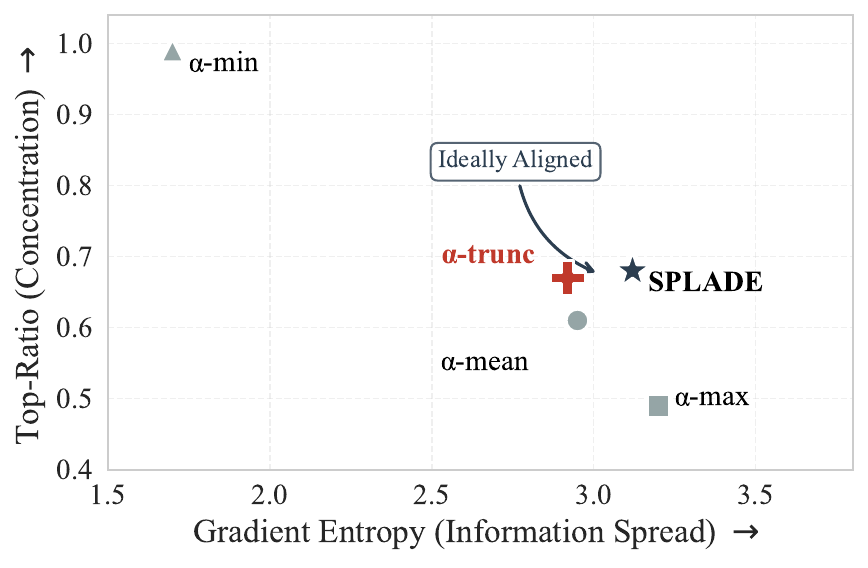} 
    \caption{Gradient analysis of truncation strategies. 
    Our \textbf{$\alpha$-trunc} (red) achieves the best alignment with SPLADE, balancing sparsity and expressiveness compared to baselines.}
    \label{fig:grad} 
\end{figure}

To provide insights into the behavior of different truncation strategies, we analyze their gradient dynamics. 
Figure~\ref{fig:grad} visualizes the trade-off between information spread (Entropy) and concentration (Top-Ratio), 
serving as a reference for understanding sparsity patterns rather than a direct measure of retrieval performance.
\definecolor{splade_noise}{RGB}{200, 50, 50} 
\definecolor{llm_insight}{RGB}{0, 100, 200}
\definecolor{common_gray}{RGB}{100, 100, 100}

The higher document-side entropy indicates broader term activation, likely from a small sparsity coefficient. However, the subsequent contrastive and FLOPS regularization gradually penalize redundant activations, producing a compact sparse index upon convergence.

\subsection{Effect of Stage 1 Training}
\label{sec:stage1_ablation}

We investigate the contribution of Stage 1 training to
retrieval effectiveness and representation sparsity.
As shown in Table~\ref{tab:stage1_ablation},
removing Stage 1 reduces MRR@10 from 0.3998 to
0.3619, indicating its importance for retrieval
effectiveness.

Beyond retrieval performance, Stage 1 also produces
substantially sparser representations.
For the same 202,827 passages, the total number of
index entries decreases from 72.10M to 42.72M,
corresponding to a 40.7\% reduction.
Similarly, the average number of non-zero terms per
passage decreases from 355.46 to 210.62.
These results demonstrate that Stage 1 improves
retrieval effectiveness while reducing index complexity,
achieving a more favorable balance between
effectiveness and sparsity.

\begin{table}[t]
\centering
\scriptsize
\caption{
Impact of Stage 1 training on retrieval effectiveness
and index sparsity.
}
\label{tab:stage1_ablation}
\setlength{\tabcolsep}{5pt}
\begin{tabular}{lcccc}
\toprule
\textbf{Training Setting}
& \textbf{\# Passages}
& \textbf{Index Size}
& \shortstack{\textbf{Avg. Terms}\\\textbf{per Passage}}
& \textbf{MRR@10} \\
\midrule
Without Stage 1 & 202,827 & 72,096,705 & 355.46 & 0.3619 \\
With Stage 1    & 202,827 & 42,719,847 & 210.62 & \textbf{0.3998} \\
\bottomrule
\end{tabular}
\end{table}
\subsection{Retrieval Performance After KL and Contrastive Fine-Tuning}
We evaluate the final model obtained after Stage 2 on MS MARCO (Dev)~\cite{nguyen2016ms} 
and DL2019/DL2020~\cite{craswell2020overview,craswell2021overviewtrec2020deep}. 
Table~\ref{tab:final_results} shows that LLM-SPLADE ($\alpha$-trunc) achieves competitive retrieval performance. 
Stage 2 joint KL adnd contrastive training leads to stable effectiveness on MS MARCO and six selected  BEIR. 
NDCG@10 on DL19 is slightly lower for short, fact-based queries relying on few key tokens. 
For example, in ``\text{tracheids are part of \_\_\_\_}'', original query tokens account for 47.5\% of total weight in SPLADE++ but 13.9\% in LLM-SPLADE, demonstrating that our model distributes weights more evenly across relevant terms while preserving core semantic contributions. 

\begin{table}[t]
\caption{
Retrieval performance on MS MARCO Dev (MRR@10)
and TREC DL19/DL20 (NDCG@10).
Best results are highlighted in bold.
$\dagger$ indicates statistically significant improvement
over SPLADE++ on MS MARCO ($p<0.01$).
}
\label{tab:final_results}
\centering
\small
\setlength{\tabcolsep}{6pt}
\begin{tabular}{lccc}
\toprule
\textbf{Method}
& \textbf{MS MARCO}
& \textbf{DL19}
& \textbf{DL20} \\
\midrule

\multicolumn{4}{l}{\textit{Dense Retrieval}} \\
ANCE~\cite{xiong2020approximate}
& 0.330 & 0.645 & 0.646 \\
ColBERTv2~\cite{santhanam2022colbertv2}
& 0.397 & \textbf{0.750} & 0.746 \\
\midrule

\multicolumn{4}{l}{\textit{Sparse Baselines}} \\
BM25~\cite{robertson2009probabilistic}
& 0.184 & 0.506 & 0.480 \\
uniCOIL~\cite{lin2021few}
& 0.351 & 0.702 & 0.674 \\
SPLADEv2~\cite{formal2021spladev2}
& 0.368 & 0.729 & -- \\
SPLADE++~\cite{formal2022distillation}
& 0.383 & 0.731 & 0.720 \\
\midrule

\multicolumn{4}{l}{\textit{LLM-based Sparse Retrieval}} \\
Lion-sp~\cite{zeng2025scaling}
& \textbf{0.410} & 0.747 & \textbf{0.751} \\
\midrule

\textbf{LLM-SPLADE (Ours)}
& 0.408$^\dagger$ & 0.721 & 0.727 \\
\bottomrule
\end{tabular}
\end{table}

\begin{table}[t]
\centering
\scriptsize
\caption{
Zero-shot NDCG@10 on six selected BEIR datasets.
Best results are highlighted in bold.
$\dagger$ indicates statistically significant
improvements over SPLADE++ ($p<0.05$).
}
\label{tab:beir_full}
\setlength{\tabcolsep}{3.8pt}
\begin{tabular}{lccc}
\toprule
\textbf{Dataset} & \textbf{SPLADE++} & \textbf{Lion-sp} & \textbf{Ours} \\
\midrule
TREC-COVID     & 72.5 & \textbf{78.3} & 74.9 \\
NFCorpus       & 34.5 & \textbf{36.5} & 35.9 \\
FiQA-2018      & 34.9 & \textbf{37.8} & 36.1 \\
ArguAna        & 51.8 & 48.8 & \textbf{55.8}$^\dagger$ \\
Touche-2020    & 24.2 & \textbf{33.0} & 27.7$^\dagger$ \\
SciFact        & 71.0 & \textbf{73.2} & 71.6 \\
\midrule
\textbf{Avg.} & 48.2 & \textbf{51.3} & 50.3 \\
\bottomrule
\end{tabular}
\end{table}

As shown in Table~\ref{tab:beir_full}, we evaluate the
zero-shot generalization of our model on six selected
BEIR datasets without additional domain-specific training.
Our method achieves an average NDCG@10 of 50.3,
outperforming SPLADE++ (48.2) and approaching
the performance of Lion-sp (51.3).
Notably, our method consistently outperforms SPLADE++
across all six datasets, with statistically significant
improvements on ArguAna and Touche-2020 ($p<0.05$).
On ArguAna, our method achieves the best NDCG@10
of 55.8, surpassing both SPLADE++ (51.8) and
Lion-sp (48.8).
Although Lion-sp achieves higher scores on most datasets,
our method maintains competitive retrieval effectiveness
across diverse out-of-domain settings.
These results suggest that our training strategy
generalizes beyond the MS MARCO training distribution.

\subsection{Qualitative Analysis}
\label{sec:qualitative_analysis}

To provide qualitative insights into the learned sparse representations, Table~\ref{table:case_study} presents a comparative analysis of expansion terms generated by the baseline SPLADE++ and our LLM-SPLADE for two representative queries.

In the first example (\textit{Ford Fusion}), the baseline appears to prioritize surface-level co-occurrences, such as \textit{crash}. In contrast, our model activates functional terms like \textit{sensor} and \textit{mechanism}. This suggests that by distilling LLM knowledge, the model is better positioned to capture the underlying technological context of "Pre-Collision Assist," going beyond mere lexical matching.

Regarding the second example (\textit{Laravel}), we observe a difference in tail distribution handling. While SPLADE++ generates some contextually ambiguous terms (e.g., \textit{tomb}, \textit{fake}), likely due to statistical artifacts, our method—guided by the $\alpha$-truncation strategy—tends to mitigate such noise. By focusing on core concepts like \textit{verification}, our approach maintains a more focused retrieval representation in this instance.

\begin{table}[!htbp]
\centering
\vspace{-2mm}
\small
\setlength{\tabcolsep}{3pt}
\renewcommand{\arraystretch}{1.4}
\begin{tabularx}{\columnwidth}{l | >{\raggedright\arraybackslash}X}
\toprule
\multicolumn{2}{l}{\textbf{Q1:} \textit{how ford fusion pre collision assist with pedestrian detection works}} \\
\midrule
\textbf{SPLADE++} & fusion (1.82), ford (1.66), pedestrian (1.60), detection (1.54), pre (1.44), collision (1.38), assist (1.22), brake (0.95), vehicle (0.91), \textbf{crash (0.83)} \\
\midrule
\textbf{Ours} & fusion (1.34), pre (1.32), ford (1.26), pedestrian (1.25), collision (1.07), detection (0.90), brake(0.75), \textcolor{llm_insight}{\textbf{sensor (0.72)}}, \textcolor{llm_insight}{\textbf{function (0.72)}}, \textcolor{llm_insight}{\textbf{mechanism (0.71)}} \\
\midrule
\midrule
\multicolumn{2}{l}{\textbf{Q2:} \textit{How does Laravel authentication work?}} \\
\midrule
\textbf{SPLADE++} & lara (2.69), \#\#vel (2.62), authentication (1.95), authentic (1.43), work (1.35), \textcolor{splade_noise}{\textit{tomb (0.63)}}, security (0.60), \textcolor{splade_noise}{\textit{button (0.54)}}, \textcolor{splade_noise}{\textit{fake (0.43)}} \\
\midrule
\textbf{Ours} & authentication (1.45), lara (1.37), work (1.17), la (1.05), authentic (1.01), \#\#vel (0.94), \textcolor{llm_insight}{\textbf{function (0.75)}}, \textcolor{llm_insight}{\textbf{mechanism (0.72)}}, \textcolor{llm_insight}{\textbf{verification (0.35)}}, security (0.36) \\
\bottomrule
\end{tabularx}
\caption{Qualitative comparison showing that our model captures better semantics (e.g., \textit{sensor}) and reduces noise (e.g., \textit{tomb}) compared to the baseline.}
\label{table:case_study}
\end{table}
\subsection{Efficiency vs. Effectiveness Trade-off}
\label{sec:efficiency}

\begin{table}[!htbp]
\centering
\footnotesize 
\caption{End-to-End efficiency comparison on MS MARCO Dev (6980 queries). Latency measures the average processing time per query (ms/q). Our method is $\mathbf{11\times}$ faster than Lion-sp while reducing storage overhead by $\mathbf{6.4\times}$.}
\label{tab:efficiency}

\setlength{\tabcolsep}{2pt} 
\begin{tabularx}{\columnwidth}{l c >{\centering\arraybackslash}X >{\centering\arraybackslash}X >{\centering\arraybackslash}X}
\toprule
\textbf{Method} & \textbf{Backbone} & \textbf{Sparsity} ($L_0$) $\downarrow$ & \textbf{Storage} (GB) $\downarrow$ & \textbf{E2E Lat.} (ms) $\downarrow$ \\
\midrule
SPLADE++ & BERT (110M) & \textbf{113} & \textbf{8} & \textbf{40.64} \\
Lion-sp  & LLaMA (1B)  & 1170 & 83 & 1046.28 \\
\midrule
\textbf{Ours} & \textbf{LLaMA (1B)} & \underline{182} & \underline{13} & \underline{89.87} \\
\bottomrule
\end{tabularx}
\end{table}

\begin{figure}[!htbp]
    \centering 
    \includegraphics[width=0.9\linewidth]{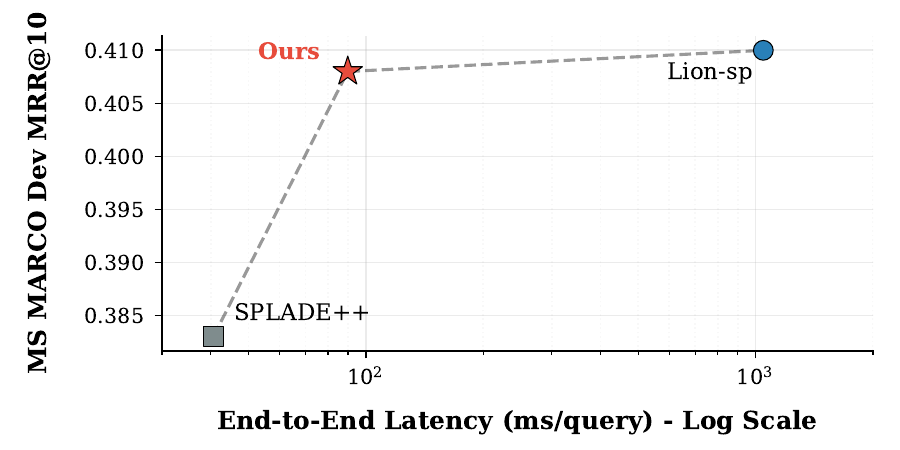} 
    \caption{Pareto comparison of latency and MRR@10 on MS MARCO Dev}
    \label{fig:parto} 
\end{figure}
Table~\ref{tab:efficiency} reports the end-to-end (E2E) efficiency on MS MARCO Dev. In this work, E2E latency refers to the online query-time cost, consisting of two components: (1) query encoding into a sparse representation via LLM inference, and (2) sparse retrieval over the inverted index using conventional term matching. This two-part decomposition is crucial for understanding the practical deployability of LLM-based sparse retrievers, as the encoding cost often dominates the overall latency in real-world settings.

While Lion-sp achieves strong effectiveness (Table~\ref{tab:final_results}), it suffers from severe inefficiency due to highly dense representations ($L_0 = 1170$), leading to substantial storage overhead (83GB) and latency exceeding 1000ms per query. Such overhead renders Lion-sp impractical for most production environments, particularly those with strict latency budgets (e.g., < 200ms) or limited GPU resources for online inference. The root cause lies in its dense activation patterns: with over a thousand non-zero terms per document, both the index footprint and the computational cost of scoring grow proportionally.

Our method addresses this limitation through a two-stage training framework that combines adaptive sparse supervision with FLOPS regularization. Specifically, dynamic $\alpha$-truncation is applied only during the alignment stage to guide the learning of compact sparse representations, without requiring additional truncation at inference time. The resulting model achieves an average document sparsity of $L_0 = 182$, reducing index storage by 6.4$\times$ (83GB $\rightarrow$ 13GB) and E2E query latency by over 11$\times$ (1046ms $\rightarrow$ $\sim$90ms) compared with Lion-sp. The substantially fewer activated terms contribute to more efficient inverted-index retrieval, while the compact index further reduces storage requirements. These results demonstrate that our training strategy achieves a favorable effectiveness--efficiency trade-off without introducing additional inference-time sparsification.
Despite adopting a larger backbone (LLaMA-1B) than SPLADE++ (BERT-110M), our approach maintains comparable efficiency, with both storage (13GB) and latency (90ms) remaining in the same order of magnitude as SPLADE++. As shown in Figure~\ref{fig:parto}, our model lies on a favorable Pareto frontier, achieving higher MRR@10 than SPLADE++ while substantially reducing latency relative to Lion-sp. This positions our framework as a practically viable solution for deploying LLM-enhanced sparse retrieval in latency-sensitive, resource-constrained environments, where both storage budget and response time are critical constraints.

\section{Conclusion}

We presented a lightweight framework for aligning LLM representations with sparse lexical retrieval through input-adaptive $\alpha$-truncation. By selecting input-specific sparse supports from SPLADE anchor representations, the proposed dynamic truncation mechanism helps alleviate head collapse and suppress long-tail activations while preserving salient semantic signals. Experimental results on MS MARCO and BEIR show that our method achieves competitive retrieval effectiveness under substantially stricter sparsity constraints. In particular, with $L_0=182$ and an end-to-end latency of approximately 90 ms per query, our model reduces index storage by 6.4$\times$ and improves query-time efficiency by over 11$\times$ compared with prior LLM-based sparse retrieval, while remaining compatible with inverted indexing.
\bibliographystyle{splncs04}
\bibliography{references}
\end{document}